# Some remarks on possible superconductivity of composition $Pb_9CuP_6O_{25}$

P. Abramian[1], A. Kuzanyan[2], V. Nikoghosyan[2], S. Teknowijoyo[3], and A. Gulian[3]*

[1] *CIEMAT Centro de Investigaciones Energéticas, Medioambientales y Tecnológicas, Madrid, Spain*
[2] *Institute for Physics Research, National Academy of Sciences, Ashtarak, Armenia*
[3] *Advanced Physics Laboratory, Institute for Quantum Studies, Chapman University, Burtonsville, MD 20866, USA*

*gulian@chapman.edu

Abstract – A material called LK-99, a modified-lead apatite crystal structure with the composition $Pb_{10-x}Cu_x(PO_4)_6O$ ($0.9 < x < 1.1$) has been reported to be an above-room-temperature superconductor at ambient pressure. It is hard to expect that it will be straightforward for other groups to reproduce the original results. We provide here some remarks which may be helpful for a success.

## 1. Introduction

A recent extraordinary claim of above-room-temperature superconductivity in $Pb_9CuP_6O_{25}$ [1-3] attracted a large amount of attention in the community. Many research groups are trying to reproduce these results, and the first reports are not yet positive [4,5]. In particular, Liu et al. [4] reproduced all the synthesis stages described in the original report; however, the resistivity vs. temperature has semiconductor-type behavior, and magnetization also increases when cooling down. The authors claim that the situation demands "more careful re-examination" which is undoubtedly true. Our remarks are devoted to this task. Here we provide experimental data on superconducting Pb films which have some parallels with the observations of Liu et al. [4], and can spread some light on the ongoing physical processes in the candidate LK-99 materials.

## 2. Experimental data

We deposited Pb films of various thickness on sapphire substrates. Details of deposition are described in Methods. For thicker films the resistivity vs. temperature looks like the traditional superconducting transition, while for thinner films, the curves are similar to those for semiconductors, Figs. 1 and 2. One should note that despite the transition taking place towards higher-values of R, its location at H=0 (Fig. 1) is at T≈7K, which coincides with the $T_c$ of Pb. Moreover, when increasing H, it moves towards lower values of T, as it should for traditional superconductivity. Magnetic measurements indeed confirm that we are dealing with a real superconducting material, Fig. 3(a).

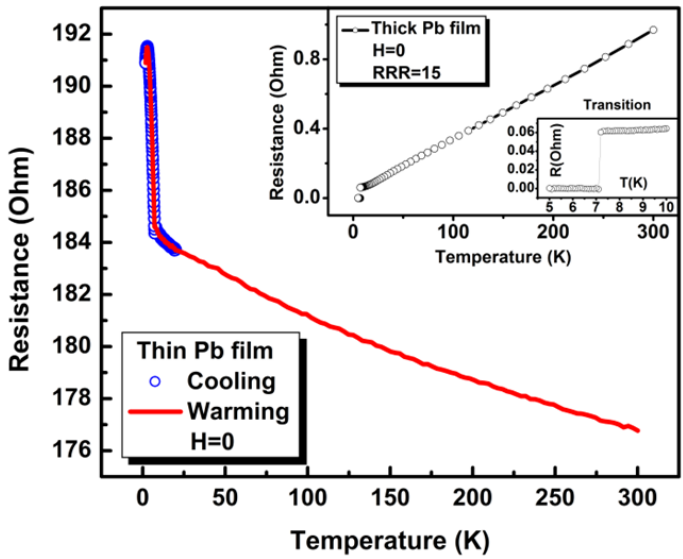

**Fig. 1.** Superconducting transition in thin and thick (inset) lead films at H=0.

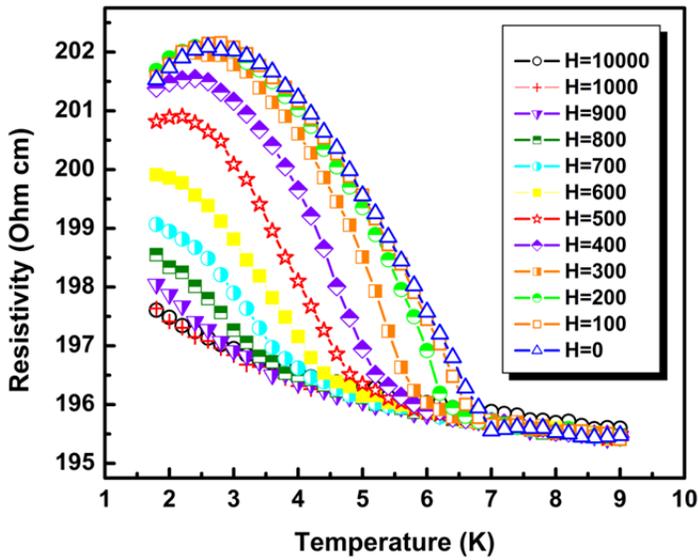

**Fig. 2.** Superconducting transition in thin Pb films at different values of the magnetic field (Oe units).

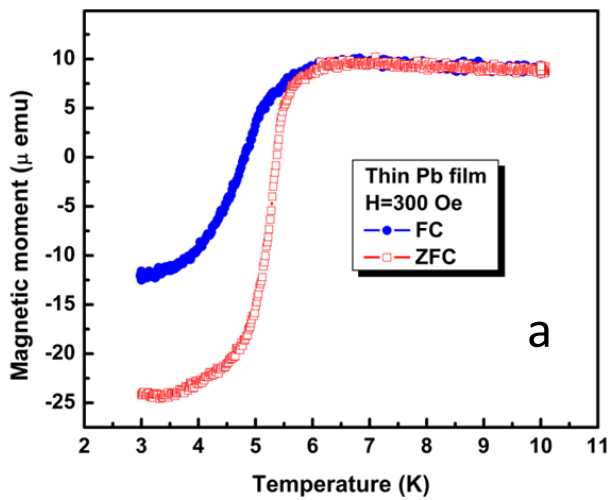
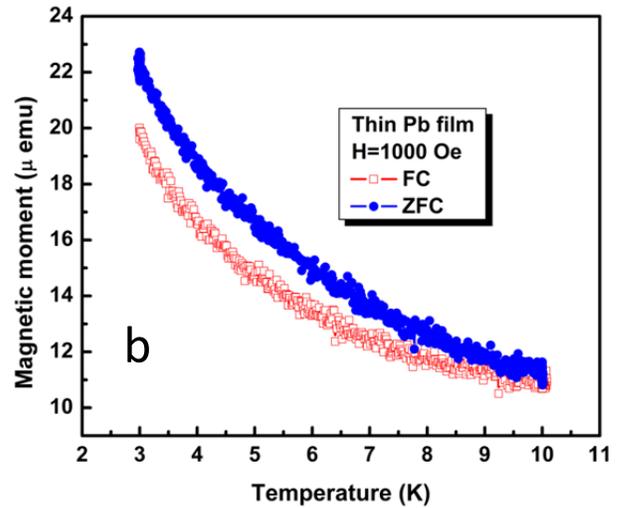

**Fig. 3.** Magnetization of thin Pb film in fields of different strength. Its resistive behavior is shown in Fig. 2.

## 3. Discussion

The curves in Fig. 3(b) have similarity with Fig. 7 of [4], and the curves in Fig. 1 and Fig. 2 with Fig. 11 of the same report. Such behaviour of Pb films, though obtained at different deposition conditions, is well-documented in scientific literature and can be explained by electron hopping from one superconducting nano-island to another one via macroscopic quantum tunneling (see, e.g., [6-9] and refs. therein). Figure 4 confirms this statement.

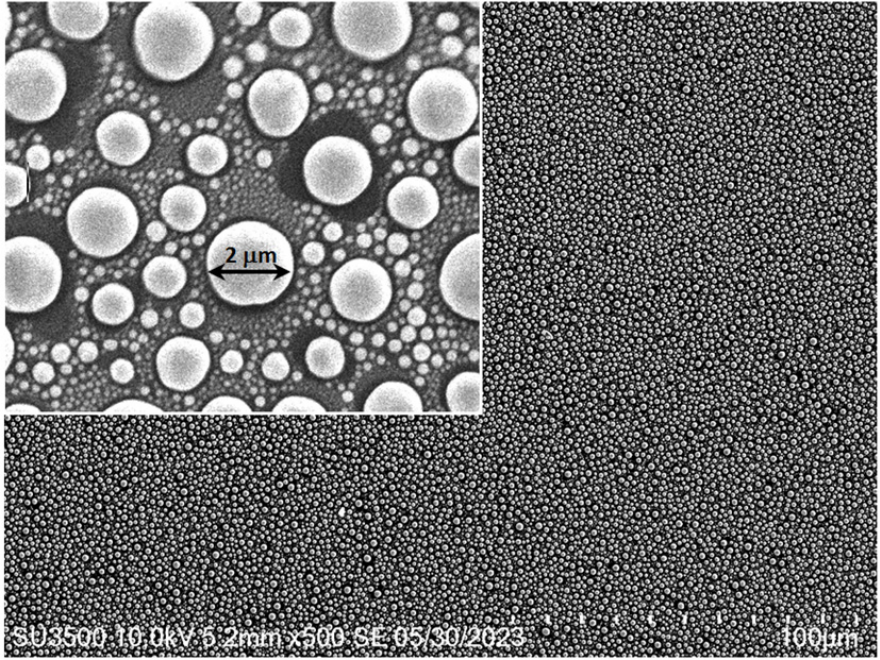

**Fig. 4.** Thin film of Pb deposited on sapphire substrate on top of Nb precursor is not wetting wetting it and a nano-island structure results with the quantum tunneling of electrons between isles, yielding the behavior documented in Figs.1-3.

This phenomenon is typical for superconductor-dielectric quantum phase transitions. It is very likely that the material which was supposed to be $Pb_{10-x}Cu_x(PO_4)_6O$ (0.9<x<1.1) is in reality a heterophase compound which contains polycrystals separated by an amorphous surrounding phase or phases. These phases do not contribute significant X-ray peaks in the registered XRD, but essentially affect the resistive behavior, and the magnetic properties as well. Indeed, any molar ratio of initial ingredients $Cu_3P$ and lanarkite $Pb_2(SO_4)O$ cannot end up in a material having Cu/P ratio 1/6 in a single-phase substance. More extensive EDX/EBSD/EDX results are required for understanding the phase content for the described route of preparation.

## 4. Conclusion

Most likely, the material LK99 as reported in [1-3] is a heterophase structure, with co-existent non-superconducting constituents. This may yield superconducting droplets surrounded by non-superconducting material and cause the phenomenon described above for the case of Pb films. Depending on the very specific details of synthesis, the degree of this effect may be stonger or weaker yieding misleading results when approaching the required composition. We agree with the conclusion expressed in [4] that the situation needs more careful re-examination and investigation. It is, indeed, an uphill task [5].

## 5. Methods

Our thin Pb films were deposited on C-cut sapphire substrates. They were heated up to $700^0$C in deposition chamber with the base vacuum $1\times10^{-8}$ Torr and gently bombarded by $Ar^+$ ions using Kauffman source for 3 min for surface preparation. Precursor Nb was deposited at $600^0$C with 1.5" DC sputtering gun during 1 min for better Pb adhesion. Then substrate was cooled down to $200^0$C, and Pb was deposited during 7 min at 10 mTorr pressure using 2" RF sputtering gun at power 150W. The cooling process, as well as other temperature variations, were performed at $30^0$C/min rate. The thick films were deposited without precursor layer. There was no substrate heating applied. Surface cleaning was performed for 5 min via ion bombardment by Kauffman source. Same RF power as for thin films was used, but for longer deposition time: 20 min. Resistive and magnetic measurements were performed on Quantum Design PPMS.

*Acknowledgements*
We are grateful to V. P. S. Awana for private correspondence and discussions. Physics Art Frontiers is acknowledged for the provided technical assistance. This research was supported by the ONR grants No. N00014-21-1-2879 and No. N00014-20-1-2442.